\documentclass[conference]{IEEEtran}
\IEEEoverridecommandlockouts
\usepackage{setspace}
\usepackage{cite}
\usepackage{amsmath,amssymb,amsfonts,mathtools,bm}
\usepackage{amsthm}
\usepackage{algorithm}
\usepackage{graphicx}
\usepackage{textcomp}
\usepackage{xcolor,float}
\usepackage{tabularx}
\usepackage{textcomp}
\usepackage{diagbox}
\usepackage{svg}
\usepackage{booktabs}
\usepackage{subfigure}
\usepackage{hyperref}

\newtheorem{problem}{Problem}
\usepackage{algpseudocode}
\usepackage{graphicx}
\usepackage{makecell}
\usepackage{amsthm}
\usepackage{caption}
\usepackage{sidecap}
\usepackage{microtype}
\usepackage{booktabs} 
\usepackage{multirow}
\usepackage{float}
\usepackage{adjustbox} 
\usepackage{amsmath}
\usepackage{amssymb}
\usepackage{colortbl}
\usepackage{makecell}
\usepackage{float}
\usepackage{url}
\usepackage{balance}
\usepackage{bbding}
\usepackage{hyperref}
\usepackage{graphicx}
\usepackage{sidecap}
\usepackage{microtype}
\usepackage{graphicx}
\usepackage{subfigure}
\usepackage{booktabs} 
\usepackage{multirow}
\usepackage{array}
\usepackage{subcaption}
\usepackage{graphicx}
\usepackage{makecell}
\usepackage{amsmath}
\usepackage{amssymb}

\usepackage{tikz}

\usepackage{changes}
\usepackage{ifthen}
\newboolean{showchanges}
\setboolean{showchanges}{false}  

\newcommand{\add}[1]{%
    \ifthenelse{\boolean{showchanges}}%
        {\textcolor{blue}{#1}}
        {#1\relax}
}

\definecolor{lime}{HTML}{A6CE39}
\DeclareRobustCommand{\orcidicon}{%
    \begin{tikzpicture}
    \draw[lime, fill=lime] (0,0) 
    circle [radius=0.16] 
    node[white] {{\fontfamily{qag}\selectfont \tiny ID}};    \draw[white, fill=white] (-0.0625,0.095) 
    circle [radius=0.007];    \end{tikzpicture}
    \hspace{-2mm}}
\foreach \x in {A, ..., Z}{%
    \expandafter\xdef\csname orcid\x\endcsname{\noexpand\href{https://orcid.org/\csname orcidauthor\x\endcsname}{\noexpand\orcidicon}}
    }

\allowdisplaybreaks
\renewcommand{\baselinestretch}{1}

\begin{document}

\title{RMSup: Physics-Informed Radio Map Super-Resolution for Compute-Enhanced Integrated Sensing and Communications}
\author{
Qiming Zhang\IEEEauthorrefmark{1},
\IEEEauthorblockN{Xiucheng Wang\IEEEauthorrefmark{2},
Nan Cheng\IEEEauthorrefmark{2},
Zhisheng Yin\IEEEauthorrefmark{2},
Xiang Li\IEEEauthorrefmark{2}
}
\IEEEauthorblockA{
\IEEEauthorrefmark{1}School of Artificial Intelligence, Xidian University, Xi'an, 710071, China\\
\IEEEauthorrefmark{2}State Key Laboratory of ISN and School of Telecommunications Engineering, Xidian University, Xi'an, 710071, China\\
Email: \{23009200991, xcwang\_1\}@stu.xidian.edu.cn, dr.nan.cheng@ieee.org, \{zsyin, lixiang\}@xidian.edu.cn}
}

    \maketitle

\IEEEdisplaynontitleabstractindextext

\IEEEpeerreviewmaketitle

\begin{abstract}
Radio maps (RMs) provide a spatially continuous description of wireless propagation, enabling cross-layer optimization and unifying communication and sensing for integrated sensing and communications (ISAC). However, constructing high-fidelity RMs at operational scales is difficult, since physics-based solvers are time-consuming and require precise scene models, while learning methods degrade under incomplete priors and sparse measurements, often smoothing away critical discontinuities. We present RMSup, a physics-informed super-resolution framework that functions with uniform sparse sampling and imperfect environment priors. RMSup extracts Helmholtz equation-informed boundary and singularity prompts from the measurements, fuses them with base-station side information and coarse scene descriptors as conditional inputs, and employs a boundary-aware dual-head network to reconstruct a high-fidelity RM and recover environmental contours jointly. Experimental results show the proposed RMsup achieves state-of-the-art performance both in RM construction and ISAC-related environment sensing.
\end{abstract}

\begin{figure*}[t] 
    \centering
    \includegraphics[width=0.9\textwidth, trim=3cm 7cm 3cm 3cm, clip]{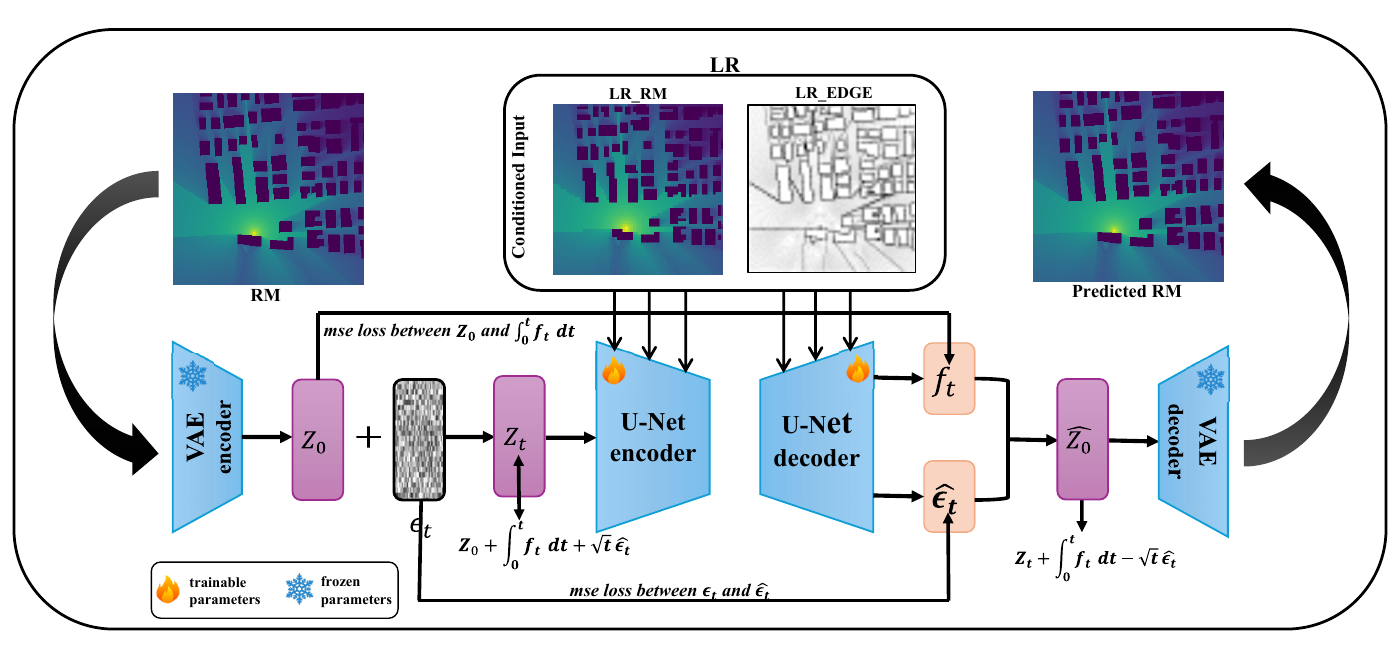}
    \caption{Training procedure of the super-resolution diffusion model.}
    \label{fig_train}
\end{figure*}

\section{Introduction}
Radio maps (RMs) offer a spatially continuous characterization of wireless propagation metrics such as path loss, received signal strength (RSS), and normalized channel power, serving as a critical enabler for integrating physical environments with network operations and service optimization \cite{zeng2024tutorial}. By elevating discrete link measurements into geometry-aware fields, RMs provide essential priors for cross-layer adaptation—facilitating rate control, power allocation, modulation selection, beamforming, mobility management, and interference coordination with scene-specific precision. Beyond conventional connectivity, RMs unify communication and sensing within a shared spatial framework, enabling tasks such as environmental reconstruction, localization, tracking, and obstacle detection for advancing integrated sensing and communications (ISAC) \cite{zeng2024tutorial}. In network deployment and operation, RMs support both pre-deployment planning and post-deployment optimization, while continuously informing digital twin systems for robust lifecycle management. As wireless systems evolve toward 6G \cite{shen2023toward}, the role of RMs becomes increasingly vital, especially under mmWave and THz propagation, ultra-dense and cell-free deployments, and reconfigurable intelligent surface–assisted environments, where compute-enhanced networking and closed-loop autonomy demand accurate, predictive spatial intelligence \cite{zeng2021toward}. However, realizing high-resolution RMs with low latency and strong generalization across diverse environments remains a fundamental challenge, which motivates the developments presented in this work.

Current methods for RM construction fall into two broad categories: physics-based electromagnetic simulation and data-driven learning.While physics-based solvers offer strong physical fidelity and interpretability, they suffer from high computational latency, often taking minutes to process environments at the scale of a floor or an entire city, and they require precisely specified inputs such as CAD geometry, material properties, and base-station parameters \cite{jones2013theory}. These methods are sensitive to unmodeled dynamics and minor inaccuracies, making them costly to maintain and unsuitable for online inference. In contrast, learning-based approaches offer fast inference and platform flexibility but degrade significantly when prior information is incomplete. Without access to accurate environmental or base-station details, these models lack electromagnetic priors, leading to excessive smoothing, hallucinated structures, and failure to capture key discontinuities such as blockage edges, diffraction boundaries, and LoS transitions \cite{jahne2005digital}. They are also highly vulnerable to domain shift and perform poorly under sparse measurements \cite{jahne2005digital}. What remains lacking is a unified framework that can leverage sparse and imperfect inputs while enforcing physically consistent constraints to recover critical boundary and singular structures. This motivates our development of a physics-informed super-resolution paradigm that reconciles physical interpretability with learning flexibility.

To overcome the limitations of existing approaches, we propose RMSup, a physics-informed super-resolution framework for radio map reconstruction under uniform sparse sampling and incomplete environmental priors. RMSup introduces Helmholtz-equation–driven boundary and singularity priors, serving as physics prompts to guide learning toward electromagnetic consistency. These priors are extracted by inverting sparse measurements through a wave-based model, yielding structural cues that highlight critical discontinuities such as LoS transitions, high-transmission boundaries, and diffraction edges \cite{jones2013theory}. The resulting boundary indicators are encoded as conditional inputs alongside uncertain base-station locations and coarse environmental descriptors. These are fused by a boundary-aware super-resolution network that jointly estimates a high-fidelity radio map and recovers geometric contours for downstream sensing. A boundary-weighted consistency loss enforces local structural accuracy, while optional PDE residual regularization enhances training stability without requiring exhaustive scene annotations \cite{jones2013theory}. RMSup achieves low-latency inference and robust generalization, preserving edge structures even under degraded priors. It supports key ISAC tasks such as localization, obstacle detection, and RIS control. Empirically, RMSup surpasses strong baselines across diverse datasets, particularly in boundary error, global consistency, and geometry recovery, demonstrating the effectiveness of physics-informed learning for compute-enhanced ISAC. The main contributions of this paper are summarized as follows.
\begin{enumerate}
    \item We propose RMSup, which fuses uniform sparse measurements with imperfect environment priors through Helmholtz-induced boundary and singularity prompts, yielding EM-consistent radio maps and recoverable scene contours for ISAC.
    \item We invert sparse samples with the Helmholtz equation to derive indicator maps of LoS transitions and diffraction edges, encode them with base-station side information as conditional channels, and employ a dual-head network for RM super-resolution and geometry recovery.
    \item Experimental results demonstrate that the proposed Helmholtz-based structural guidance consistently enhances radio map reconstruction accuracy and spatial fidelity across both diffusion-based and CNN-based architectures.
\end{enumerate}

\begin{figure}[h]
    \centering
    \includegraphics[width=0.4\textwidth]{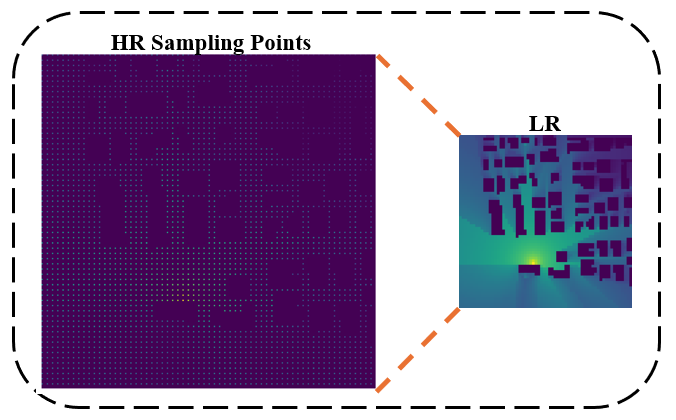}
    \caption{Uniform sampling of the high-resolution RM illustrating evenly spaced measurement points.}
    \label{fig_sampling}
    \vspace{-4mm}  
\end{figure}


\section{Preliminary}

\subsection{Score-Based Denoising Diffusion Model}
Diffusion models are a class of generative models that can produce high-quality samples from complex data distributions \cite{ho2020denoising}. Score-based diffusion models implement a generative process via stochastic differential equations (SDEs), where data is gradually corrupted and then reconstructed by learning the gradient of the log-probability density, called the score function. Unlike traditional diffusion models that use discrete Markov chains, score-based models perturb data continuously in time, which makes them well-suited for inverse problems such as radio map reconstruction, while naturally enabling efficient Bayesian sampling. The forward process is given by:
\begin{align}
d\bm{x} = f(\bm{x}, t) dt + g(t) d\bm{w},
\end{align}
where $\bm{x}$ is the data variable, $f(\bm{x}, t)$ governs its deterministic evolution, $g(t)$ scales stochastic perturbations, and $d\bm{w}$ is a standard Wiener process. As time progresses from $0$ to $T$, the distribution $p_t(\bm{x})$ gradually approaches an isotropic Gaussian, allowing tractable sampling.  

The reverse-time generative process is described by:
\begin{align}
d\bm{x} = \left[ f(\bm{x}, t) - g^2(t) \nabla_{\bm{x}} \log p_t(\bm{x}) \right] dt + g(t) d\bm{\bar{w}},
\end{align}
where $\nabla_{\bm{x}} \log p_t(\bm{x})$ is the score function and $d\bar{\bm{w}}$ a reverse-time Wiener process. Since $p_t(\bm{x})$ is unknown, a neural network $s_\theta(\bm{x}, t)$ is trained to approximate the score:
\begin{align}
s_\theta(\bm{x}, t) \approx \nabla_{\bm{x}} \log p_t(\bm{x}).
\end{align}
For deterministic sample generation, the equivalent probability flow ODE is:
\begin{align}
d\bm{x} = \left[ f(\bm{x}, t) - \frac{1}{2} g^2(t) \nabla_{\bm{x}} \log p_t(\bm{x}) \right] dt,
\end{align}
which allows generating samples without stochasticity, similar to denoising diffusion probabilistic models (DDPMs). In DDPMs, the forward process is discretized as:
\begin{align}
q(\bm{x}_t \mid \bm{x}_{t-1}) = \mathcal{N}(\bm{x}_t; \alpha_t \bm{x}_{t-1}, \beta_t \bm{I}),
\end{align}
and the score function is approximated by a learned denoiser $\epsilon_\theta$:
\begin{align}
s_\theta(\bm{x}, t) = -\frac{\epsilon_\theta(\bm{x}, t)}{\sqrt{1 - \bar{\alpha}_t}}.
\end{align}
This shows that DDPM can be viewed as a discrete, variance-preserving approximation of continuous score-based diffusion models.  

The decoupled diffusion model (DDM) separates data attenuation and noise injection into two stages \cite{huang2024decoupled}. Unlike conventional diffusion models that directly add Gaussian noise, DDM first attenuates the initial state $\bm{n}_0$ toward zero, then injects stochastic noise. The forward process is a continuous Markov process with Gaussian transitions:
\begin{align}
q\left(\bm{x}_t \mid \bm{x}_0\right) = \mathcal{N}\left(\gamma_t \bm{x}_0, \delta_t^2 \bm{I}\right),
\end{align}
where $\gamma_t$ and $\delta_t$ control attenuation and noise variance. Its differential form is:
\begin{align}
d \bm{x}_t &= f_t \bm{x}_t dt + g_t d\bm{\epsilon}_t,\\
f_t &= \frac{d \log \gamma_t}{dt},\\
g_t^2 &= \frac{d \delta_t^2}{dt} - 2 f_t \delta_t^2,
\end{align}
with $f_t$ governing decay and $g_t^2$ the accumulation of noise, which stabilizes the diffusion early on. The reverse process reconstructs $\bm{x}_0$:
\begin{align}
d \bm{x}_t = \left[f_t \bm{x}_t - g_t^2 \nabla_{\bm{x}} \log q\left(\bm{x}_t\right)\right] dt + g_t d\overline{\bm{\epsilon}}_t.
\end{align}
The deterministic transformation to zero further simplifies the forward process:
\begin{align}
q(\bm{x}_t|\bm{x}_0) = \mathcal{N}\left(\bm{x}_0 + \int_0^t \bm{f}_t \, dt,\, t \bm{I}\right),
\end{align}
enabling efficient reverse sampling:
\begin{align}
q\left(\bm{x}_{t-\Delta t} \mid \bm{x}_t, \bm{x}_0\right) &= \mathcal{N}\Bigg(\bm{x}_t + \int_t^{t-\Delta t} \bm{f}_t \, dt \notag\\
&\quad - \frac{\Delta t}{\sqrt{t}} \bm{\epsilon}_t,\, \frac{\Delta t(t-\Delta t)}{t} \bm{I}\Bigg).
\end{align}

\subsection{Helmholtz Wave Equation}
The Helmholtz equation can be derived from Maxwell’s equations in a source-free, linear, isotropic, and time-invariant medium. Considering time-harmonic fields with an $e^{j\omega t}$ dependence, where $\omega$ is the angular frequency, the curl equations are:
\begin{align}
\nabla \times \bm{E} = -j\omega\mu \bm{G}, \quad \nabla \times \bm{G} = j\omega\epsilon \bm{E},
\end{align}
where $\bm{E}$ and $\bm{G}$ denote the electric and magnetic fields, and $\epsilon$ and $\mu$ are the permittivity and permeability of the medium. Taking the curl of Faraday's law and substituting Ampère's law yields:
\begin{align}
\nabla \times (\nabla \times \bm{E}) = -j\omega\mu (j\omega\epsilon \bm{E}) = -\omega^2 \mu\epsilon \bm{E}.
\end{align}
Using the vector identity $\nabla \times (\nabla \times \bm{E}) = \nabla(\nabla \cdot \bm{E}) - \nabla^2 \bm{E}$ and assuming a source-free medium ($\nabla \cdot \bm{E} = 0$), the equation reduces to
\begin{align}
\nabla^2 \bm{E} + \omega^2 \mu \epsilon \bm{E} = 0.
\end{align}
Defining the wavenumber $k = \omega \sqrt{\mu \epsilon}$, we obtain the vector Helmholtz equation:
\begin{align}
\nabla^2 \bm{E} + k^2 \bm{E} = 0.\label{raw-helmholtz}
\end{align}
For fields with a radiation source, the Helmholtz equation is written as
\begin{align}
\nabla^2 \bm{E} + k^2 \bm{E} + \bm{e} = 0,
\end{align}
where $\bm{e}$ represents the source term.

\section{System Model and Problem Formulation}

We consider the construction of a high-resolution RM over a discrete two-dimensional spatial region represented by an $N \times N$ grid. Each grid cell is assumed sufficiently small such that the pathloss within it remains approximately constant. The RM can thus be expressed as a matrix $\bm{P} \in \mathbb{R}^{N \times N}$, where each element $P(i,j)$ denotes the pathloss at the corresponding grid location.

Obtaining a full high-resolution RM is often costly due to measurement limitations. A low-resolution radio map $\bm{P}_{\mathrm{LR}} \in \mathbb{R}^{M \times M}$ $(M < N)$ is first acquired via uniform or sparse sampling. Additionally, an edge map $\bm{K}$ highlighting regions of rapid power variation could in principle be estimated from measurements. However, low-resolution measurements rarely contain sufficient detail to reliably extract edge information. To provide structural guidance, we construct the low-resolution edge map $\bm{K}_{\mathrm{LR}} \in \mathbb{R}^{M \times M}$ by computing edges from the high-resolution map $\bm{P}$ and downsampling. This idealized edge map allows us to evaluate the performance upper bound of the super-resolution model.

The objective is to reconstruct a high-resolution RM $\hat{\bm{P}} \in \mathbb{R}^{N \times N}$ that approximates the ground truth $\bm{P}$ given $\bm{P}_{\mathrm{LR}}$ and $\bm{K}_{\mathrm{LR}}$. A neural network $\bm{\mu}_{\bm{\theta}}(\cdot)$ parameterized by $\bm{\theta}$ is designed to infer the fine-grained pathloss distribution. The reconstruction is trained to minimize the discrepancy with the ground truth under a loss function $\mathcal{L}(\hat{\bm{P}}, \bm{P})$, leading to the following optimization problem:
\begin{problem}\label{p1_modified}
    \begin{align}
        &\min_{\bm{\theta}} && \mathcal{L}(\hat{\bm{P}}, \bm{P}), \label{obj_modified} \\
        &\text{s.t.} && \hat{\bm{P}} = \bm{\mu}_{\bm{\theta}}(\bm{P}_{\mathrm{LR}}, \bm{K}_{\mathrm{LR}}). \tag{\ref{obj_modified}a}
    \end{align}
\end{problem}

\section{Physics-Informed Super-Resolution Framework for Radio Map Reconstruction}

\subsection{Discrete Helmholtz Curvature for Edge Extraction}

Given a RM representing the EM power $I = |u|^2$, we first compute the amplitude
\begin{equation}
A = \sqrt{I}.
\end{equation}

The Laplacian of $A$ is discretized using a standard 5-point finite-difference scheme:
\begin{align}
\frac{\partial^2 A}{\partial x^2}\bigg|_{(i,j)} \!\approx\! \frac{A(i{+}1,j)-2A(i,j)+A(i{-}1,j)}{h^2},\\
\frac{\partial^2 A}{\partial y^2}\bigg|_{(i,j)} \!\approx\! \frac{A(i,j{+}1)-2A(i,j)+A(i,j{-}1)}{h^2},
\end{align}
\begin{align}
\nabla_h^2 A(i,j)\approx(&A(i{+}1,j)+A(i{-}1,j)+A(i,j{+}1)\\
&+A(i,j{-}1)-4A(i,j))/h^2,
\end{align}
and the following curvature indicators are computed \cite{wang2025radiodiffk2helmholtzequationinformed}:
\begin{align}
k_{\text{eff}}^2(i,j) &= -\frac{\nabla_h^2 A(i,j)}{A(i,j)+\varepsilon},\\
k_{\log}(i,j) &= -\nabla_h^2\log\!\big(A(i,j)+\varepsilon\big),
\end{align}
where $\varepsilon$ is a small constant to ensure numerical stability. These indicators characterize local spatial variations of the EM power and are robust to mild smoothing and gain variations.

Finally, a binary edge map, also referred to as the K map, is obtained by thresholding the effective curvature:
\begin{equation}
K(i,j) =
\begin{cases}
1, & k_{\text{eff}}^2(i,j) < 0,\\
0, & k_{\text{eff}}^2(i,j) \ge 0.
\end{cases}
\end{equation}

This binary map highlights regions of rapid power variation, effectively capturing the high-curvature zones in the radio map. It serves as a structural prior to guide the training of super-resolution models, providing physics-informed information about potential electromagnetic singularities.


\subsection{Low-Resolution RM and Edge Map Construction}


\textit{\textbf{Low-Resolution RM Construction: }}
To simulate a realistic sparse measurement scenario, the high-resolution radio map 
$P \in \mathbb{R}^{N\times N}$ is uniformly sampled on a regular grid with interval $s$ 
in both spatial dimensions (as illustrated in Figure~\ref{fig_sampling}), yielding a low-resolution map 
$P_{LR} \in \mathbb{R}^{n\times n}$, where $n = N/s$. 
Mathematically, this process can be expressed as
\begin{align}
P_{LR}(i,j) &= P(s\,i,\, s\,j), \quad i,j = 1,\dots,n,
\end{align}
which corresponds to selecting every $s$-th pixel along both axes. 
This operation effectively reduces spatial resolution while maintaining the overall propagation structure.
Before being fed into the network, $P_{LR}$ is linearly normalized to $[0,1]$ for numerical stability.

\textit{\textbf{Low-Resolution Edge Map Construction: }}
Following \cite{wang2025radiodiffk2helmholtzequationinformed}, an edge-enhanced map \(K \in \mathbb{R}^{N\times N}\) is derived from the RM \(P\) to emphasize rapid spatial variations in received power, which correlate with propagation discontinuities and building boundaries. For each low-resolution pixel \((i,j)\),
\begin{equation}
K_{LR}(i,j)=\sum_{u=0}^{1}\sum_{v=0}^{1} w_{uv},K!\left(p_u,q_v\right),
\quad i,j=1,\ldots,n,
\tag{1}
\end{equation}
where \((p_0,q_0),(p_0,q_1),(p_1,q_0),(p_1,q_1)\) denote the four neighboring coordinates on the reference grid. The bilinear weights are given by
\begin{equation}
w_{uv}=\bigl(1-\lvert x_i-p_u\rvert\bigr)\bigl(1-\lvert y_j-q_v\rvert\bigr),
\tag{2}
\end{equation}
with \((x_i,y_j)\) the continuous location of pixel \((i,j)\) in the same coordinate system. This resampling preserves the dominant spatial structure encoded in \(K\) while enforcing geometric consistency with \(P_{LR}\). Prior to network ingestion, \(K_{LR}\) is min–max normalized to \([0,1]\) to improve numerical stability. Used as a physics-informed auxiliary channel, \(K_{LR}\) supplies structural priors that accentuate physically plausible discontinuities and suppress artifacts. Looking ahead, this formulation can be extended to incorporate additional priors—e.g., environmental geometry, propagation constraints, and scene semantics—to enhance robustness and cross-scene generalization.

\begin{figure*}[t]
\centering
\captionsetup{font={small}, skip=6pt}

\begin{tabular}{ccccc}
\includegraphics[width=0.15\linewidth]{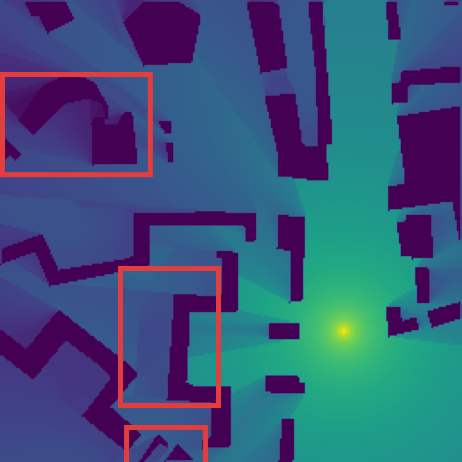} &
\includegraphics[width=0.15\linewidth]{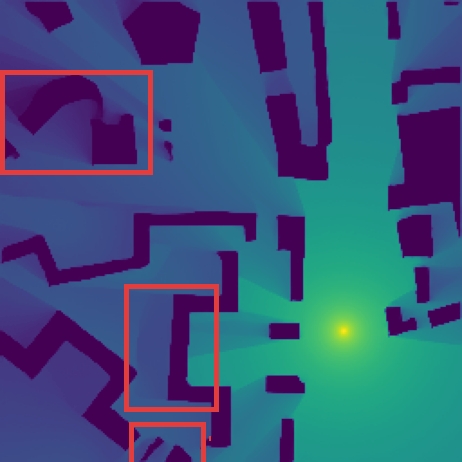} &
\includegraphics[width=0.15\linewidth]{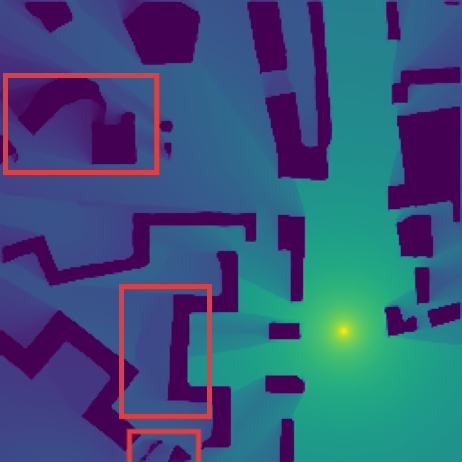} &
\includegraphics[width=0.15\linewidth]{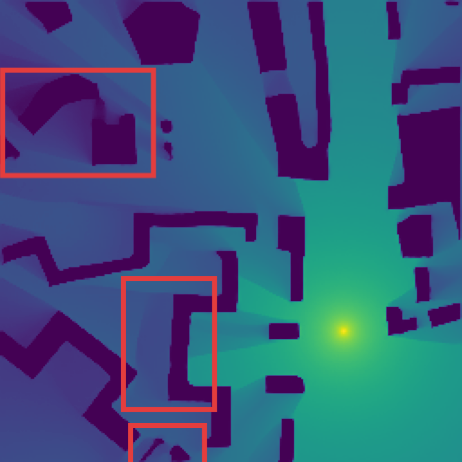} &
\includegraphics[width=0.15\linewidth]{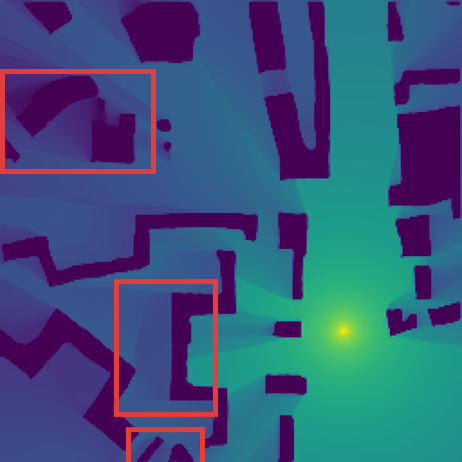} \\

\includegraphics[width=0.15\linewidth]{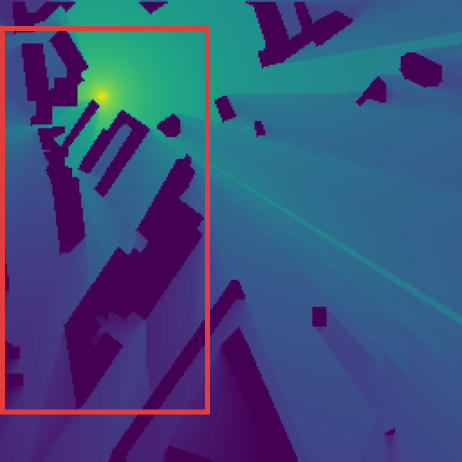} &
\includegraphics[width=0.15\linewidth]{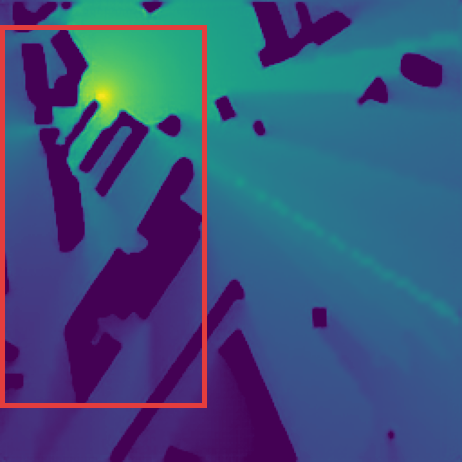} &
\includegraphics[width=0.15\linewidth]{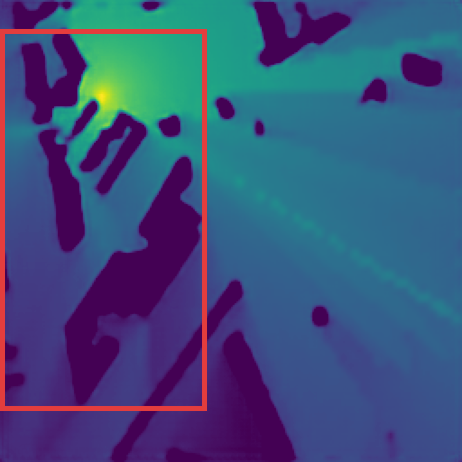} &
\includegraphics[width=0.15\linewidth]{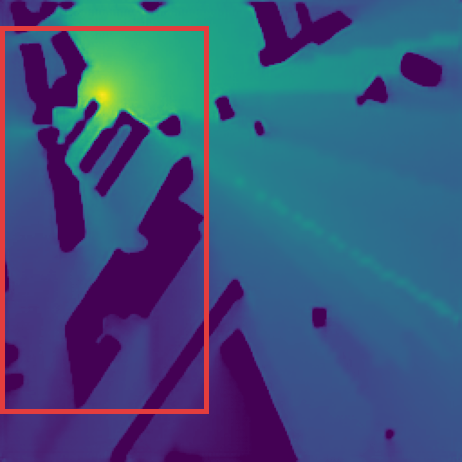} &
\includegraphics[width=0.15\linewidth]{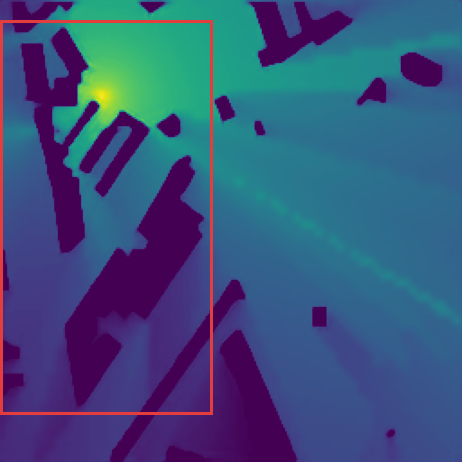} \\

\raisebox{0pt}{Ground Truth} &
\raisebox{0pt}{$K$-edge} &
\raisebox{0pt}{LBP} &
\raisebox{0pt}{Canny} &
\raisebox{0pt}{Base} 
\end{tabular}

\vspace{-5pt}
\caption{Qualitative comparison of different edge-guided methods. 
The first row shows RMs reconstructed by \textbf{RadioDiff} under various guidance strategies, 
while the second row presents the corresponding results generated by \textbf{RadioUNet}.}
\label{fig:five_images_adjusted}
\end{figure*}

\subsection{Physics-Informed Super-Resolution Framework}

For the super-resolution task, the network input is constructed by concatenating the low-resolution radio map 
$P_{LR} \in \mathbb{R}^{n\times n}$ with its corresponding edge map 
$K_{LR} \in \mathbb{R}^{n\times n}$. 
This combined input provides both amplitude and structural priors, enabling the network to reconstruct the high-resolution radio map 
$\hat{P} \in \mathbb{R}^{N\times N\times 1}$ with an upscaling factor of $s = N/n$. 

The RMSup model is trained in a two-stage physics-informed framework combining latent-space distribution learning and conditional diffusion refinement. In the first stage, a Variational Autoencoder (VAE) models the intrinsic distribution of high-resolution radio maps. Each map $\bm{x}$ is encoded into a latent representation $\bm{z}$, approximating the posterior $q_\phi(\bm{z}|\bm{x})$, while the decoder reconstructs $\bm{x}$ via $p_\theta(\bm{x}|\bm{z})$. The training minimizes the negative evidence lower bound (ELBO), which includes a reconstruction term and a Kullback–Leibler (KL) divergence:
{\small
\begin{align}
\mathcal{L}_{\mathrm{VAE}} 
= \mathbb{E}_{q_\phi(\bm{z}|\bm{x})} \big[ -\log p_\theta(\bm{x}|\bm{z}) \big] 
+ D_{\mathrm{KL}}\big( q_\phi(\bm{z}|\bm{x}) \| p(\bm{z}) \big),
\end{align}
}
where the reconstruction loss uses the $\ell_2$-norm between the original and reconstructed maps~\cite{kingma2019introduction}. This stage captures global structure and produces a compact, physics-consistent latent representation for the refinement stage.

In the second stage, a conditional denoising UNet operates in the latent space to predict the deterministic drift $\hat{\bm{f}}_t$ and stochastic noise $\hat{\bm{\epsilon}}_t$ of the diffusion process, conditioned on the concatenated input $(P_{LR}, K_{LR})$. The corresponding losses are
\begin{align}
\mathcal{L}_{\text{drift}} = \mathbb{E}_t \big[ \| \hat{\bm{f}}_t + \bm{z}_0 \|_2^2 \big], \quad
\mathcal{L}_{\text{noise}} = \mathbb{E}_t \big[ \| \hat{\bm{\epsilon}}_t - \bm{\epsilon}_t \|_2^2 \big].
\end{align}
An auxiliary constraint enforces latent consistency between the predicted and ground-truth latent variables:
\begin{align}
\mathcal{L}_{\text{recon}} = \| \hat{\bm{z}}_0 - \bm{z}_0 \|_2^2.
\end{align}
The total loss combines all components:
\begin{align}
\mathcal{L}_{\text{total}} = \lambda_1 \mathcal{L}_{\text{drift}} + \lambda_2 \mathcal{L}_{\text{noise}} + \lambda_3 \mathcal{L}_{\text{recon}},
\end{align}
where $\lambda_1$, $\lambda_2$, and $\lambda_3$ balance the contributions of each term.

By jointly optimizing the VAE and conditional diffusion modules, RMSup reconstructs high-resolution RM from low-resolution inputs while preserving both amplitude accuracy and structural sharpness. The framework ensures physics-consistent super-resolution guided by latent-space regularization and edge-based structural priors, as illustrated in Figure.~\ref{fig_train}.

\section{Experimental}
\subsection{Experimental Setup}

To evaluate the proposed method, experiments are conducted on the Static Radio Maps (SRM) subset of RadioMapSeer \cite{DatasetPaper}, which contains 700 urban maps with varying numbers of buildings derived from OpenStreetMap data. Each map includes 80 transmitter positions and corresponding pathloss ground truth (GT), represented as $256 \times 256$ binary morphological images, where a pixel value of 1 indicates a building interior and 0 indicates an open area. Transmitters and receivers are at 1.5 m height, buildings at 25 m, transmit power 23 dBm, and carrier frequency 5.9 GHz. SRM is generated using the Dominant Path Model (DPM) \cite{dpm}, considering only the main propagation path and large-scale static buildings. The GTs are obtained via electromagnetic ray tracing, serving as a benchmark for comparing neural network methods with computational EM methods.

The training set includes 600 environments with 80 base-station placements each, and evaluation is performed on 100 unseen environments, also with 80 base-station placements each. This zero-shot, cross-environment setup allows assessment of the model's generalization to novel layouts and varying building densities, with high-resolution maps downsampled by a factor of 4 from $256 \times 256$ to $64 \times 64$ for network input.

\subsection{Quantitative Analysis of Image Metrics}

To evaluate the effectiveness of different structural guidance strategies in radio map super-resolution, we design four configurations for both model types:

\begin{itemize}
    \item \textbf{$K$-edge \cite{wang2025radiodiffk2helmholtzequationinformed}:} Our proposed guidance based on discrete Helmholtz curvature, designed to highlight electromagnetic propagation boundaries and capture geometry-aware structural information.

    \item \textbf{LBP \cite{OJALA199651}:} Local Binary Pattern texture mapping that encodes local structural textures. 
    Although it captures surface roughness, it lacks the ability to represent global spatial structure.

    \item \textbf{Canny \cite{canny1986computational}:} The classical Canny edge detector using Gaussian smoothing and gradient-based edge extraction. 
    
    \item \textbf{Base:} A baseline configuration without structural guidance, relying solely on the low-resolution power map input.
\end{itemize}

To verify the generality of $K$-edge guidance, we employ two representative architectures:

\begin{itemize}
    \item \textbf{RadioUNet} \cite{levie2021RadioUNet}: A sampling-free CNN-based method for RM reconstruction. It employs a U-Net trained with supervised learning to infer RMs from environmental information and serves as a foundational benchmark due to its simplicity and effectiveness.

    \item \textbf{RadioDiff} \cite{wang2024radiodiff}: The current state-of-the-art in sampling-free RM construction. It formulates the task as a conditional generative problem based on a DDM, combining a VAE and denoising UNet to model reverse-time EM propagation in latent space for accurate and high-fidelity reconstruction.
\end{itemize}


From Table~\ref{tab:sr_results}, we observe that the proposed $K$-edge guidance consistently improves super-resolution performance across both RadioDiff and RadioUnet models. For RadioDiff, $K$-edge achieves the lowest RMSE (0.0229) and NMSE (0.0061), as well as the highest SSIM (0.9718) and PSNR (33.15 dB) among all configurations. Compared with the Base setting, the improvements are substantial, demonstrating that the inclusion of geometry-aware structural information allows the diffusion model to better capture high-frequency details and structural consistency in the reconstructed radio maps. Similarly, for RadioUnet, $K$-edge attains the lowest RMSE (0.0292) and NMSE (0.0097), and the highest SSIM (0.9437) and PSNR (30.93 dB). While traditional guidance like LBP or Canny provides some gains over the Base configuration, $K$-edge consistently outperforms them, highlighting the importance of physically consistent structural cues.

\begin{table}[h]
\captionsetup{font={small}, skip=6pt}
\centering
\caption{\textbf{Quantitative Comparison of Diffusion and UNet methods on SRM.} Bold red and underlined blue indicate the highest and second highest values for each method. The corresponding visual reconstructions are shown in Figure~\ref{fig:five_images_adjusted}}
\vspace{-6pt}
\resizebox{0.95\linewidth}{!}{
\begin{tabular}{@{}cc|cccc@{}}
\toprule
\multicolumn{2}{c|}{Method} & RMSE $\downarrow$ & NMSE $\downarrow$ & SSIM $\uparrow$ & PSNR $\uparrow$ \\ \midrule
\multirow{4}{*}{RadioDiff} 
& $K$-edge & {\color[HTML]{9A0000}\textbf{0.0229}} & {\color[HTML]{9A0000}\textbf{0.0061}} & {\color[HTML]{9A0000}\textbf{0.9718}} & {\color[HTML]{9A0000}\textbf{33.15}} \\
& LBP & {\color[HTML]{00009B}\underline{0.0244}} & {\color[HTML]{00009B}\underline{0.0067}} & {\color[HTML]{00009B}\underline{0.9657}} & {\color[HTML]{00009B}\underline{32.50}} \\
& Canny & 0.0245 & 0.0070 & 0.9658 & 32.52 \\
& Base & 0.0505 & 0.0298 & 0.8948 & 26.23 \\ \midrule
\multirow{4}{*}{RadioUnet} 
& $K$-edge & {\color[HTML]{9A0000}\textbf{0.0292}} & {\color[HTML]{9A0000}\textbf{0.0097}} & {\color[HTML]{9A0000}\textbf{0.9437}} & {\color[HTML]{9A0000}\textbf{30.93}} \\
& LBP & 0.0348 & 0.0136 & 0.9173 & 29.32 \\
& Canny & {\color[HTML]{00009B}\underline{0.0315}} & {\color[HTML]{00009B}\underline{0.0112}} & {\color[HTML]{00009B}\underline{0.9357}} & {\color[HTML]{00009B}\underline{30.25}} \\
& Base & 0.0527 & 0.0321 & 0.8773 & 25.81 \\ \bottomrule
\end{tabular}
}
\label{tab:sr_results}
\end{table}


\subsection{Spatial Awareness Evaluation}

To assess the spatial reconstruction capability of different structural guidance strategies, we compute the Intersection over Union (IOU) between the predicted radio maps and the ground-truth building layouts. 
During downsampling of the SRM dataset, uniform step-size sampling results in substantial loss of fine-grained building information. 
Thus, this evaluation focuses on how well each guidance method can recover the structural boundaries of buildings.

For IOU computation, we first binarize both the predicted and ground-truth maps, treating pixels with values below a threshold (set to 10) as background. 
The IOU is then calculated as:
\begin{align}
\text{IOU} = \frac{|P \cap G|}{|P \cup G|}
\end{align}
where $P$ and $G$ denote the predicted and ground-truth binary building masks, respectively.

\begin{table}[h]
\captionsetup{font={small}, skip=6pt}
\centering
\caption{\textbf{IOU comparison for RadioDiff and RadioUnet methods on SRM.} Bold red and underlined blue indicate the highest and second highest values for each method.}
\vspace{-6pt}
\resizebox{0.75\linewidth}{!}{
\begin{tabular}{@{}c|cccc@{}}
\toprule
Method & $K$-edge & LBP & Canny & Base \\ \midrule
RadioDiff & {\color[HTML]{9A0000}\textbf{0.9609}} & {\color[HTML]{00009B}\underline{0.9591}} & 0.9577 & 0.8822 \\
RadioUnet & {\color[HTML]{9A0000}\textbf{0.9328}} & 0.8934 & {\color[HTML]{00009B}\underline{0.9220}} & 0.8670 \\ \bottomrule
\end{tabular}
}
\label{tab:sr_iou_compact}
\end{table}

Table~\ref{tab:sr_iou_compact} summarizes the IOU results for RadioDiff and RadioUnet with different guidance inputs. 
We observe that $K$-edge consistently achieves the highest IOU values for both models, indicating that it better preserves and reconstructs building structures. 
Traditional guidance methods like LBP or Canny provide moderate improvements over the Base configuration, but do not match the geometry-aware performance of $K$-edge. 
These results validate the advantage of $K$-edge guidance in enhancing spatially aware super-resolution, particularly for recovering fine structural details in urban environments.

\section{Conclusion}

In this work, we proposed RMSup, a physics-informed super-resolution framework for radio map reconstruction. By integrating low-resolution measurements with Helmholtz-guided structural priors, RMSup effectively preserves high-frequency details and geometric features. Extensive experiments on urban radio map datasets demonstrate that our method achieves superior reconstruction accuracy, spatial consistency, and generalization compared to baseline models, highlighting the benefits of combining physical knowledge with learning-based approaches for integrated sensing and communications.

\bibliography{ref}

@article{zeng2021toward,
  title={Toward environment-aware {6G} communications via channel knowledge map},
  author={Zeng, Yong and Xu, Xiaoli},
  journal={IEEE Wireless Commun.},
  volume={28},
  number={3},
  pages={84--91},
  year={2021}
}

@article{ho2020denoising,
  title={Denoising diffusion probabilistic models},
  author={Ho, Jonathan and Jain, Ajay and Abbeel, Pieter},
  journal={Advances in neural information processing systems},
  volume={33},
  pages={6840--6851},
  year={2020}
}

@article{shen2023toward,
  title={Toward immersive communications in {6G}},
  author={Shen, Xuemin and Gao, Jie and Li, Mushu and Zhou, Conghao and Hu, Shisheng and He, Mingcheng and Zhuang, Weihua},
  journal={Frontiers in Computer Science},
  volume={4},
  pages={1068478},
  year={2023},
  publisher={Frontiers Media SA}
}

@article{zeng2024tutorial,
  title={A tutorial on environment-aware communications via channel knowledge map for {6G}},
  author={Zeng, Yong and Chen, Junting and Xu, Jie and Wu, Di and Xu, Xiaoli and Jin, Shi and Gao, Xiqi and Gesbert, David and Cui, Shuguang and Zhang, Rui},
  journal={{IEEE} Commun. Surveys Tuts.},
  year={2024},
  volume={26},
  number={3},
  pages={1478-1519},
  publisher={IEEE}
}

@article{VAE,
  title={Auto-encoding variational bayes},
  author={Kingma, Diederik P and Welling, Max},
  journal={arXiv preprint arXiv:1312.6114},
  year={2013}
}

@article{kingma2019introduction,
  title={An introduction to variational autoencoders},
  author={Kingma, Diederik P and Welling, Max and others},
  journal={Foundations and Trends{\textregistered} in Machine Learning},
  volume={12},
  number={4},
  pages={307--392},
  year={2019},
  publisher={Now Publishers, Inc.}
}

@book{jahne2005digital,
  title={Digital image processing},
  author={J{\"a}hne, Bernd},
  year={2005},
  publisher={Springer Science \& Business Media}
}

@inproceedings{dpm,
  title={Dominant path prediction model for urban scenarios},
  author={Wahl, Ren{\'e} and W{\"o}lfle, Gerd and Wertz, Philipp and Wildbolz, Pascal and Landstorfer, Friedrich},
  booktitle={Proceedings of 14th IST mobile and wireless communications summit},
  pages={1--5},
  year={2005}
}

@misc{huang2024decoupled,
      title={Decoupled Diffusion Models: Simultaneous Image to Zero and Zero to Noise}, 
      author={Yuhang Huang and Zheng Qin and Xinwang Liu and Kai Xu},
      year={2024},
      eprint={2306.13720},
      archivePrefix={arXiv},
      primaryClass={cs.CV}
}

@book{jones2013theory,
  title={The theory of electromagnetism},
  author={Jones, Douglas Samuel},
  year={2013},
  publisher={Elsevier}
}

@article{levie2021radiounet,
  title={{RadioUNet}: Fast radio map estimation with convolutional neural networks},
  author={Levie, Ron and Yapar, {\c{C}}a{\u{g}}kan and Kutyniok, Gitta and Caire, Giuseppe},
  journal={ {IEEE} Trans. Wireless Commun.},
  volume={20},
  number={6},
  pages={4001--4015},
  year={2021},
  publisher={IEEE}
}

@article{wang2024radiodiff,
  title={{RadioDiff}: An Effective Generative Diffusion Model for Sampling-Free Dynamic Radio Map Construction},
  author={Wang, Xiucheng and Tao, Keda and Cheng, Nan and Yin, Zhisheng and Li, Zan and Zhang, Yuan and Shen, Xuemin},
  journal={IEEE Trans. Cognit. Commun. Networking, Early access},
  year={2024},
  pages={1-13},
  publisher={IEEE}
}

@ARTICLE{canny1986computational,
  author={Canny, John},
  journal={IEEE Transactions on Pattern Analysis and Machine Intelligence},
  title={A Computational Approach to Edge Detection},
  year={1986},
  volume={PAMI-8},
  number={6},
  pages={679-698},
  doi={10.1109/TPAMI.1986.4767851}
}

@article{OJALA199651, title = {A comparative study of texture measures with classification based on featured distributions}, journal = {Pattern Recognition}, volume = {29}, number = {1}, pages = {51-59}, year = {1996}, issn = {0031-3203}, doi = {https://doi.org/10.1016/0031-3203(95)00067-4}, url = {https://www.sciencedirect.com/science/article/pii/0031320395000674}, author = {Timo Ojala and Matti Pietikäinen and David Harwood} }

@misc{wang2025radiodiffk2helmholtzequationinformed,
  title={RadioDiff-$k^2$: Helmholtz Equation Informed Generative Diffusion Model for Multi-Path Aware Radio Map Construction}, 
  author={Xiucheng Wang and Qiming Zhang and Nan Cheng and Ruijin Sun and Zan Li and Shuguang Cui and Xuemin Shen},
  year={2025},
  eprint={2504.15623},
  archivePrefix={arXiv},
  primaryClass={cs.LG},
  url={https://arxiv.org/abs/2504.15623}
}

@article{DatasetPaper,
  url = {https://arxiv.org/abs/2212.11777},
  journal={arXiv preprint:2212.11777},
  author = {Yapar, {\c{C}}a{\u{g}}kan and Levie, Ron and Kutyniok, Gitta and Caire, Giuseppe},
  title = {Dataset of Pathloss and {ToA} Radio Maps With Localization Application},
  publisher = {arXiv},
  year = {2022}
}
\bibliographystyle{IEEEtran}
\ifCLASSOPTIONcaptionsoff
  \newpage
\fi
\end{document}